\documentclass[11pt]{article}
\usepackage{amssymb}

\begin{document}

\title{ Electroweak corrections to the decays of top squark and gluino } \vspace{3mm}

\author{{ Hou Hong-Sheng$^{1}$, Ma Wen-Gan$^{1,2}$,Wan Lang-Hui$^{1}$ and Zhang Ren-You$^{1}$}\\
{\small $^{1}$Department of Modern Physics, University of Science and Technology of China (USTC), Hefei, Anhui 230027,}\\
{\small  People's Republic of China}\\
{\small $^{2}$CCAST (World Laboratory), P.O.Box 8730, Beijing
100080,P.R.China} }

\date{}
\maketitle
\vskip 12mm

\begin{abstract}
The electroweak corrections at the order of
$O(\alpha_{ew}m_{t(b)}^2/m_W^2)$ to the partial widths of the
$\tilde{t}_2 \rightarrow \tilde{g} + t$ and $\tilde{g} \rightarrow
\tilde{t}_1 + \bar{t} $ decays ( depending on the masses of the
particles involved ) are investigated within the supersymmetric
standard model. The relative corrections can reach the value of
$10\%$ in some parameter space, which can be comparable with the
corresponding QCD corrections. Therefore, they should be taken
into account for the precise experimental measurement at future
colliders.
\end{abstract}

\vskip 5cm


\vfill \eject

\baselineskip=0.32in

\renewcommand{\thesection}{}
\newcommand{\nb}{\nonumber}
\newcommand{\mch}{m_{\tilde{\chi}^+_i}}
\newcommand{\mno}{m_{\tilde{\chi}^0_j}}

\makeatletter      % '@' is now a normal "letter" for TeX
\@addtoreset{equation}{section}
\makeatother       % '@' is restored as a "non-letter" character for TeX

%\section{Introduction}
\begin{flushleft} {\bf 1. INTRODUCTION} \end{flushleft}
\par
Over the past few years, much effort has been devoted to searching for
new physics beyond the standard model (SM) \cite{s1}\cite{s2}. The
minimal supersymmetric model(MSSM) \cite{s3} is considered one
of the most attractive extensions of the SM. Generally in the
MSSM, two coloured scalar quarks (squarks) $\tilde q_{L}$ and
$\tilde q_{R}$, which correspond to chiral eigenstates, are required
as partners corresponding to the chiral quarks appearing in SM,
because any realistic model must be extended from SM. The physical
mass eigenstates $\tilde{q}_1$ and $\tilde{q}_2$ are the mixtures
of these chiral eigenstates. Since in general the mixing size is
proportional to the mass of the related ordinary quark \cite{sa},
the mass splitting of the physical top squarks $\tilde{t}_1$ and
$\tilde{t}_2$ can be quite large. In fact, it is likely that the
lighter top squark mass eigenstate is the lightest scalar quark in
supersymmetric theory and the mass $m_{\tilde{t}_1}$ may even be
smaller than the top mass $m_t$ itself \cite{Ellis}. This implies
that there are quite different decay scenarios in the top squark-top quark
sector depending on the mass values of the particles involved.
Therefore, the study of the scalar top quarks is of particular
interest.
\par
If the gluinos are heavy enough, scalar quarks will mainly decay
into quarks plus charginos (or neutralinos) and lighter squarks plus
vector bosons (or Higgs bosons). These decays have been
extensively discussed in the Born approximation
\cite{Baer}\cite{Bartl}. The QCD corrections of the reaction
$\tilde{q} \to q + \chi $ has been discussed in
Ref.\cite{Djouadi}, and its inverse processes: $t \to \tilde{t} +
\chi^0 $ and $t \to \tilde{b} + \chi^+$ have been discussed in
Refs.\cite{Junger}. The Yukawa corrections and the full one-loop
electroweak(EW) radiative corrections to the squark decays into
quarks plus charginos (or neutralinos) also were give in
Refs.\cite{Guasch1}\cite{Guasch2}. The QCD corrections and the
Yukawa corrections to the heavier squark decays into lighter
squarks plus vector bosons (or Higgs bosons) have been calculated
in Refs.\cite{Bartl1}\cite{Jin}.
\par
If top squark particles are heavy enough ($m_{\tilde{t}_{j}}>m_t +
m_{\tilde{g}} $ ), the decay pattern $\tilde{t}_j \to t +
\tilde{g}$ will be kinematically allowed. The theoretical
predictions of the supersymmetric (SUSY) QCD corrections for the decay channels:
$\tilde{q} \to \tilde{g}+q$ (for
$m_{\tilde{q}}>m_{\tilde{g}}+m_q$), $\tilde{g}\to
\tilde{q}+\bar{q}/\bar{\tilde{q}}+q $ (for
$m_{\tilde{g}}>m_{\tilde{q}}+m_q $) have been calculated in
Ref.\cite{Beenakker1}, where $q$ denotes quarks except the top quark.
And in Ref.\cite{Beenakker2}, the processes $\tilde{t}_j
\rightarrow \tilde{g} + t$ and $\tilde{g} \rightarrow \tilde{t}_j
+ t $ are discussed including SUSY QCD corrections.
\par
Squarks and gluinos not heavier than a few hundred $GeV$, can be
produced in significant number at high-energy hadron colliders,
i.e. the Fermilab Tevatron $p\bar{p}$ collider and the CERN Largh Hadron Collider (LHC)
in the future \cite{Beenakker3}\cite{Beenakker4}. In
Ref.\cite{Kane} it was argued that half of the top quark at
Tevatron might come from gluino decays into top and top squark,
$\tilde{g}\to t+\tilde{t}_1$. So the accurate calculations
including quantum corrections for these decays are necessary. The
largest radiative corrections for the squark-gluino sector in the
MSSM are associated with the strong interaction, the relative
SUSY-QCD corrections for the process $\tilde{t}_2 \to \tilde{g}+t$
can reach $35\%$ and that for $\tilde{g} \to \tilde{t}_1 +t$ can
reach $-10\%$\cite{Beenakker2}. However, the investigation of the
electroweak corrections for these processes is necessary in
precise experimental measurement, since their electroweak
corrections maybe also sizable and could not be neglected. In this
paper, we present the calculations of the Yukawa corrections of
the order $O(\alpha_{ew}m_{t(b)}^2/m_W^2)$ to the width of
\begin{eqnarray}
\tilde{t}_2 \rightarrow \tilde{g} + t ~~~~~~\left(m_{\tilde{t}_2}>m_t+m_{\tilde{g}}\right)\\
\tilde{g}\rightarrow \bar{t}+\tilde{t}_1~~~~~ and~~~ c.c.
~~~~~~\left(m_{\tilde{g}}>m_t+m_{\tilde{t}_1}      \right)
\end{eqnarray}
These corrections are mainly induced by Yukawa couplings from
Higgs-quark-quark couplings, Higgs-squark-squark couplings,
Higgs-Higgs-squark-squark couplings, chargino
(neutralino)-quark-squark couplings, and
squark-squark-squark-squark couplings.

\begin{flushleft} {\bf 2. Notation and tree-level result  } \end{flushleft}
\par
We summarize our notations and present the relevant interaction
Lagrangian of the MSSM as below in order to make our paper
self-contained.
\par
The tree-level top squark squared-mass matrix is written as:
\begin{eqnarray}
{\cal M}^2 &=& \left( \begin{array}{cc} {\cal M}_{LL}^2 & {\cal
M}_{LR}^2 \\ {\cal M}_{RL}^2 & {\cal M}_{RR}^2
\end{array}\right) \nb\\
&=& \left( \begin{array}{cc} M_{\tilde{Q}}^2 + m_t^2 + m_Z^2 \cos
2\beta \left(\frac{1}{2} - \frac{2}{3}s_W^2\right) &  -m_t
\left(A_t + \mu \cot \beta\right)
\\ -m_t \left(A_t + \mu \cot \beta\right) &
M_{\tilde{U}}^2+m_t^2+\frac{2}{3}m_Z^2\cos 2\beta s_W^2
\end{array}\right)
\end{eqnarray}
The parameters $M_{\tilde{Q}}, M_{\tilde{U}}$, $\mu$ and $A_t$ are
the soft-SUSY-breaking masses, SUSY Higgs mass parameter and
trilinear coupling, $m_Z$ and $s_W$ are the Z-boson mass and the
weak mixing angle, and tan$\beta$ is the ratio of the two vacuum
expectation values in the Higgs sector. The diagonal entries of
the top squark mass matrix correspond to the $L$ and $R$ squark-mass
terms, and the off-diagonal entries are due to chirality-flip Yukawa
interactions. The chiral states $\tilde{t}_L$ and $\tilde{t}_R$
are rotated into the mass eigenstates $\tilde{t}_{10}$ and
$\tilde{t}_{20}$:
\begin{equation}
\left( \begin{array}{cc} \tilde{t}_{10} \\ \tilde{t}_{20}
\end{array}
\right) = R^{\tilde{t}}\left( \begin{array}{cc} \tilde{t}_L \\
\tilde{t}_R \end{array} \right),~~R^{\tilde{t}} = \left(
\begin{array}{cc}\cos\theta_0 &
 \sin\theta_0 \\-\sin\theta_0 & \cos\theta_0 \end{array}
 \right)
\end{equation}
by these Yukawa interactions. The mass eigenvalues and the
rotation angle can be calculated from the mass matrix in Eq.(3):
\begin{eqnarray}
 m_{\tilde{t}_{1}}^2,~m_{\tilde{t}_{2}}^2&=&\frac{1}{2}\left[ {\cal
 M}_{LL}^2+{\cal M}_{RR}^2\mp\left[ ({\cal
 M}_{LL}^2-{\cal M}_{RR}^2)^2+4({\cal
 M}_{LR}^2)^2\right]^{1/2}\right]
 \\
 sin(2\theta_0)&=&\frac{2{\cal
 M}_{LR}^2}{m_{\tilde{t}_{1}}^2-m_{\tilde{t}_{2}}^2},~~~~cos(2\theta_0)=\frac{({\cal
 M}_{LL}^2-{\cal
 M}_{RR}^2)}{m_{\tilde{t}_{1}}^2-m_{\tilde{t}_{2}}^2},
\end{eqnarray}
where we assume $\tilde{t}_{1}$ to be the lighter top squark state.
\par
We can write the relevant Lagrangian density in the
($\tilde{t}_L,\tilde{t}_R$) basis as following form ($a,j ,k$ are
colour indices):
\begin{eqnarray}
 {\cal L} = -\sqrt{2}\hat{g}_s
 T_{jk}^a\left(\bar{\tilde{g}}_a P_L t^k \tilde{t}_L^{j*}+
 \bar{t}^j P_R \tilde{g}_a \tilde{t}_L^k - \bar{\tilde{g}}_a P_R
 t^k \tilde{t}_R^{j*} - \bar{t}^j P_L \tilde{g}_a \tilde{t}_R^{k}
  \right)
\end{eqnarray}
Here the $q\tilde{q}\tilde{g}$ Yukawa coupling $\hat{g}_s$ should
coincide with the $qqg$ gauge coupling $g_s$ by supersymmetry.
\par
The tree-level partial widths for the top squark and gluino decays,
$\tilde{t}_{1,2} \rightarrow \tilde{g} + t$ and $\tilde{g}
\rightarrow \tilde{t}_{1,2} + t $ as shown in Fig.1(a.1)-1(a.3),
are given by \cite{Beenakker2}:
\begin{eqnarray}
 \Gamma_{0}(\tilde{t}_{1,2} \rightarrow \tilde{g}t)&=&
 \frac{2\alpha_s\kappa}{3m_{\tilde{t}_{1,2}}^3} \left[
 m_{\tilde{t}_{1,2}}^2-m_t^2-m_{\tilde{g}}^2 \pm 2m_tm_{\tilde{g}}
 sin(2\theta_0) \right]
 \\
 \Gamma_{0}(\tilde{g}\rightarrow \bar{t}\tilde{t}_{1,2})&=&
 -\frac{\alpha_s\kappa}{8m_{\tilde{g}}^3} \left[
 m_{\tilde{t}_{1,2}}^2-m_t^2-m_{\tilde{g}}^2 \pm 2m_tm_{\tilde{g}}
 sin(2\theta_0) \right]
\end{eqnarray}
where
\begin{eqnarray}
\kappa=\left( \sum_i m_i^4-\sum_{i\neq j}m_i^2m_j^2 \right)^{1/2}
\end{eqnarray}
The sums run over all particles involved in the decay process.

\begin{flushleft} {\bf 3. Yukawa corrections  } \end{flushleft}
\par
The Feynman diagrams of the one-loop Yukawa corrections to the
processes (1) and (2) are shown in Figs.1(b.1)-(b.4). In the
calculation, we use the 't Hooft gauge and adopt the dimensional
reduction (DR) scheme\cite{Copper}, which is commonly used in the
calculation of the electroweak correction in frame of the MSSM as
it preserves supersymmetry at least at one-loop order, to control
the ultraviolet divergences in the virtual loop corrections. The
complete on-mass-shell scheme \cite{denner}\cite{hollik} is used
in doing renormalization.
\par
The relevant renormalization constants are defined as
\begin{eqnarray}
t_0 &=& \left( 1+\frac{1}{2}\delta Z_{tt}^LP_L+\frac{1}{2}\delta Z
_{tt}^RP_R \right)t
\nb \\
\tilde{t}_{i0} &=& \left( 1+\delta Z_i^{\tilde{t}}
\right)^{1/2}\tilde{t}_i+\delta Z_{ij}^{\tilde{t}}\tilde{t}_j \nb \\
\theta_0 &=& \theta+\delta\theta
\end{eqnarray}
In the on-mass-shell scheme the renormalization constants defined
in Eq.(11) can be fixed by the renormalization
conditions\cite{denner}\cite{hollik} as:
\begin{eqnarray}
\delta Z^L_{tt}&=&-\tilde{Re}\left[\Sigma^L_{tt}(m_t^2) +
m_t^2(\Sigma^{L'}_{tt}(m_t^2)+\Sigma^{R'}_{tt}(m_t^2)) +
                         m_t (\Sigma^{SL'}_{tt}(m_t^2) + \Sigma^{SR'}_{tt}(m_t^2))\right], \nb \\
\delta Z^R_{tt}&=&-\tilde{Re}\left[\Sigma^R_{tt}(m_t^2) +
m_t^2(\Sigma^{L'}_{tt}(m_t^2)+\Sigma^{R'}_{tt}(m_t^2)) +
                         m_t (\Sigma^{SL'}_{tt}(m_t^2) + \Sigma^{SR'}_{tt}(m_t^2))\right]
\end{eqnarray}
where $\Sigma{'}(m_t^2)=\frac{\partial\Sigma(p^2)}{\partial
p^2}|_{p^2=m_t^2}$. Notice that the top squark wave-function
renormalization involves the mixed scalar field renormalization
constants:
\begin{eqnarray}
\delta Z_i^{\tilde{t}} &=& -\tilde{Re}
\Sigma_{ii}^{\tilde{t}'}(m_{\tilde{t}_i}^2), \nb \\
\delta Z_{ij}^{\tilde{t}} &=&
\frac{\Sigma_{ij}^{\tilde{t}}(m_{\tilde{t}_j}^2)}{m_{\tilde{t}_j}^2-m_{\tilde{t}_i}^2}~~~~~~~~~~~(i\neq
j )
\end{eqnarray}
\par
For the counterterm of the top squark mixing angle $\theta$, using the same
renormalization scheme as Ref.\cite{Guasch1}, we can get
\begin{eqnarray}
\delta\theta=\frac{1}{2}(\delta Z_{12}^{\tilde{t}}-\delta
Z_{21}^{\tilde{t}})=\frac{1}{2}
\frac{\Sigma_{12}^{\tilde{t}}(m_{\tilde{t}_1}^2)+\Sigma_{12}^{\tilde{t}}(m_{\tilde{t}_2}^2)}
{m_{\tilde{t}_2}^2-m_{\tilde{t}_1}^2}
\end{eqnarray}
\par
Taking into account the Yukawa corrections, the renormalized
amplitude for these process is given by
\begin{eqnarray}
M_{ren}=M_{0}+\delta M
\end{eqnarray}
\par
Correspondingly, the renormalized decay width is then given by
\begin{eqnarray}
\Gamma_{ren}=\Gamma_{0}+\delta \Gamma
\end{eqnarray}
In the following we present the explicit calculation of the decay
width of $\tilde{t}_2 \to t\tilde{g}$, while the calulation for
the decays of $\tilde{t}_1\to t\tilde{g}$ and
$\tilde{g}\to\bar{t}\tilde{t}_j$ are very similar with that for
$\tilde{t}_2 \to t\tilde{g}$, so we will not present them. Now we
present the explicit formulae of the calculation of the process
$\tilde{t}_2 \to t\tilde{g}$.
\par
We denote this process as
\begin{eqnarray}
\tilde{t}_2(p_1,c_1) \to t(k_1,c_2) + \tilde{g}(k_2,a)
\end{eqnarray}
where $c_1,c_2,a$ are color indices. By substituting Eq.(11) into
the bare Lagrangian, we can obtain its counterterm.
\begin{eqnarray}
\delta {\cal L}_{\tilde{t}_2t\tilde{g}} &\equiv&-\sqrt{2}\hat{g}_s
 T_{c_1c_2}^{a} \bar{t}~\left[\delta C_L P_L+\delta C_R
 P_R\right]~\tilde{g}\tilde{t}_2 \nb\\
&=&-\sqrt{2}\hat{g}_s T_{c_1c_2}^{a} \bar{t}~\left[\left(-\delta
Z_{12}^{\tilde{t}} \sin\theta-\frac{1}{2}\delta
Z_{tt}^{R*}\cos\theta-\frac{1}{2}\delta Z_2^{\tilde{t}}
\cos\theta+\delta\theta \sin\theta\right)R_L\right.\nb\\
&+&\left.\left(\delta Z_{12}^{\tilde{t}}
\cos\theta-\frac{1}{2}\delta Z_{tt}^{L*}
\sin\theta-\frac{1}{2}\delta Z_2^{\tilde{t}}
\sin\theta-\delta\theta \cos\theta\right)P_R\right]
\end{eqnarray}
\par
Then the renormalized one-loop part of the amplitude for the decay
$\tilde{t}_2\to t+\tilde{g}$ can be written as
\begin{eqnarray}
\delta {\cal M}&=&\delta {\cal M}^c+\delta {\cal M}^v,
\end{eqnarray}
where $\delta {\cal M}^v$ and $\delta {\cal M}^c$ are contributed
by the vertex corrections and the counterterm, respectively. They
have the expressions as
\begin{eqnarray}
\delta {\cal M}^c &=&-i\sqrt{2}\hat{g}_s
 T_{c_1c_2}^{a}\bar{t}~\left[\delta C_L P_L+\delta C_R
 P_R\right]~\tilde{g}\tilde{t}_2 \nb\\
\delta {\cal M}^v &=&T_{c_1c_2}^{a}\bar{t}~\left[\Lambda_1 +
\Lambda_2 \gamma_5 \right]~\tilde{g}\tilde{t}_2
\end{eqnarray}
$\delta C_{L,R}$ are defined in Eq.(18), and $\Lambda_{1,2}$ are
the form factors contributed by the diagrams in Fig.1(b).
\par
We decompose the form factors $\Lambda_{1,2}$ in the form as
\begin{eqnarray}
\Lambda_{1,2}=\Lambda_{1,2}^{(1)}+\Lambda_{1,2}^{(2)}+
\Lambda_{1,2}^{(3)}+\Lambda_{1,2}^{(4)},
\end{eqnarray}
where $\Lambda_{1,2}^{(i)}~(i=1,2,3,4)$ are the form factors
contributed by the diagrams in Fig.1(b.1), Fig.1(b.2), Fig.1(b.3)
and Fig.1(b.4), respectively.
\begin{itemize}
\item For diagram Fig.1(b.1), we introduce the following notation:
\begin{eqnarray}
F_a^{(1)}=
V_{\tilde{g}\tilde{b}_{\beta}b}^{(1)*}V_{t\tilde{b}_{\beta}\tilde{\chi}_k^+}^{(1)}V_{b\tilde{t}_2\tilde{\chi}_k^+}^{(1)}
,~~~ F_b^{(1)}=
V_{\tilde{g}\tilde{b}_{\beta}b}^{(2)*}V_{t\tilde{b}_{\beta}\tilde{\chi}_k^+}^{(1)}V_{b\tilde{t}_2\tilde{\chi}_k^+}^{(1)}
,\nb\\
F_c^{(1)} =
V_{\tilde{g}\tilde{b}_{\beta}b}^{(1)*}V_{t\tilde{b}_{\beta}\tilde{\chi}_k^+}^{(2)}V_{b\tilde{t}_2\tilde{\chi}_k^+}^{(1)}
,~~~ F_d^{(1)}=
V_{\tilde{g}\tilde{b}_{\beta}b}^{(2)*}V_{t\tilde{b}_{\beta}\tilde{\chi}_k^+}^{(2)}V_{b\tilde{t}_2\tilde{\chi}_k^+}^{(1)}
,\nb\\
F_e^{(1)} =
V_{\tilde{g}\tilde{b}_{\beta}b}^{(1)*}V_{t\tilde{b}_{\beta}\tilde{\chi}_k^+}^{(1)}V_{b\tilde{t}_2\tilde{\chi}_k^+}^{(2)}
,~~~ F_f^{(1)}=
V_{\tilde{g}\tilde{b}_{\beta}b}^{(2)*}V_{t\tilde{b}_{\beta}\tilde{\chi}_k^+}^{(1)}V_{b\tilde{t}_2\tilde{\chi}_k^+}^{(2)}
,\nb\\
F_g^{(1)} =
V_{\tilde{g}\tilde{b}_{\beta}b}^{(1)*}V_{t\tilde{b}_{\beta}\tilde{\chi}_k^+}^{(2)}V_{b\tilde{t}_2\tilde{\chi}_k^+}^{(2)}
,~~~ F_h^{(1)}=
V_{\tilde{g}\tilde{b}_{\beta}b}^{(2)*}V_{t\tilde{b}_{\beta}\tilde{\chi}_k^+}^{(2)}V_{b\tilde{t}_2\tilde{\chi}_k^+}^{(2)}
.
\end{eqnarray}
The form factors $\Lambda_{1,2}^{(1)}$ contributed by diagram
Fig.1(b.1) are:
\begin{eqnarray}
\Lambda_1^{(1)} &=&\frac{1}{32\pi^2}\sum_{k,\beta=1}^2\left[
\left( m_t m_{\tilde{\chi}^+_k}( F_a^{(1)} +F_h^{(1)})+ (m_t^2  +
    m_t m_{\tilde{g}}) (F_c^{(1)} +F_f^{(1)})+ m_t m_b (F_d^{(1)} +
    F_e^{(1)})\right)C_{11}^{(1)}
   \right.\nb\\&+&\left((m_{\tilde{g}}-m_t) m_{\tilde{\chi}^+_k} (F_a^{(1)}+F_h^{(1)})
    + (m_{\tilde{g}}^2-m_t^2)( F_c^{(1)} +F_f^{(1)})+
    (m_{\tilde{g}}-m_t) m_b( F_d^{(1)} +F_e^{(1)})
    \right)C_{12}^{(1)}\nb\\&+& \left(
m_{\tilde{g}} m_{\tilde{\chi}^+_k}(F_a^{(1)}+F_h^{(1)})
     + m_{\tilde{\chi}^+_k} m_b (F_b^{(1)}+F_g^{(1)}) +
   ( m_t m_{\tilde{g}} + m_{\tilde{b}_{\beta}}^2) (F_c^{(1)}+F_f^{(1)})
    \right. \nb \\&+& \left.m_t m_b (F_d^{(1)}+F_e^{(1)})
   \right)C_0^{(1)}
    -\left. \left(F_c^{(1)}+F_f^{(1)}\right)B_0^{(1)}\right] \nb \\
\Lambda_2^{(1)}&=&\frac{1}{32\pi^2}\sum_{k,\beta=1}^2\left[\left(m_t
m_{\tilde{\chi}^+_k}( F_a^{(1)} -F_h^{(1)})+ (m_t^2  -
    m_t m_{\tilde{g}} )(F_c^{(1)}-F_f^{(1)}) + m_t m_b
    (F_e^{(1)}-F_d^{(1)})\right)C_{11}^{(1)}\right.
 \nb \\&+& \left((m_t  +
    m_{\tilde{g}} )m_{\tilde{\chi}^+_k}( F_h^{(1)}-F_a^{(1)})+ (m_t^2 -
    m_{\tilde{g}}^2) (F_f^{(1)}-F_c^{(1)})+(m_t  + m_{\tilde{g}}) m_b (F_d^{(1)} -
    F_e^{(1)})\right)C_{12}^{(1)}\nb\\&+&\left(m_{\tilde{g}}
m_{\tilde{\chi}^+_k} (F_h^{(1)}-F_a^{(1)}) +
    m_{\tilde{\chi}^+_k} m_b( F_g^{(1)}-F_b^{(1)}) +( m_t m_{\tilde{g}}
    -  m_{\tilde{b}_{\beta}}^2)(F_f^{(1)}-F_c^{(1)})\right.\nb\\ &+& \left. m_t m_b(F_e^{(1)}- F_d^{(1)})
    \right)C_0^{(1)} +\left.\left(F_f^{(1)}-F_c^{(1)}\right)B_0^{(1)}
    \right]
\end{eqnarray}
with $B_0^{(1)} = B_0(-p1, m_{\tilde{\chi}_k^+},m_b)$,
$C_{0,11,12}^{(1)}=C_{0,11,12}(k_1, -p_1, m_{\tilde{b}_{\beta}},
m_{\tilde{\chi}_k^+}, m_b).$
\item For diagram Fig.1(b.2), we introduce the following notation:
\begin{eqnarray}
F_a^{(2)}=
V_{\tilde{g}\tilde{t}_{\alpha}t}^{(1)*}V_{t\tilde{t}_2\tilde{\chi}_l^0}^{(1)}V_{t\tilde{t}_{\alpha}\tilde{\chi}_l^0}^{(1)}
,~~~ F_b^{(2)}=
V_{\tilde{g}\tilde{t}_{\alpha}t}^{(2)*}V_{t\tilde{t}_2\tilde{\chi}_l^0}^{(1)}V_{t\tilde{t}_{\alpha}\tilde{\chi}_l^0}^{(1)}
,\nb\\
F_c^{(2)} =
V_{\tilde{g}\tilde{t}_{\alpha}t}^{(1)*}V_{t\tilde{t}_2\tilde{\chi}_l^0}^{(2)}V_{t\tilde{t}_{\alpha}\tilde{\chi}_l^0}^{(1)}
,~~~ F_d^{(2)}=
V_{\tilde{g}\tilde{t}_{\alpha}t}^{(2)*}V_{t\tilde{t}_2\tilde{\chi}_l^0}^{(2)}V_{t\tilde{t}_{\alpha}\tilde{\chi}_l^0}^{(1)}
,\nb\\
F_e^{(2)} =
V_{\tilde{g}\tilde{t}_{\alpha}t}^{(1)*}V_{t\tilde{t}_2\tilde{\chi}_l^0}^{(1)}V_{t\tilde{t}_{\alpha}\tilde{\chi}_l^0}^{(2)}
,~~~ F_f^{(2)}=
V_{\tilde{g}\tilde{t}_{\alpha}t}^{(2)*}V_{t\tilde{t}_2\tilde{\chi}_l^0}^{(1)}V_{t\tilde{t}_{\alpha}\tilde{\chi}_l^0}^{(2)}
,\nb\\
F_g^{(2)} =
V_{\tilde{g}\tilde{t}_{\alpha}t}^{(1)*}V_{t\tilde{t}_2\tilde{\chi}_l^0}^{(2)}V_{t\tilde{t}_{\alpha}\tilde{\chi}_l^0}^{(2)}
,~~~ F_h^{(1)}=
V_{\tilde{g}\tilde{t}_{\alpha}t}^{(2)*}V_{t\tilde{t}_2\tilde{\chi}_l^0}^{(2)}V_{t\tilde{t}_{\alpha}\tilde{\chi}_l^0}^{(2)}
.
\end{eqnarray}
The form factors $\Lambda_{1,2}^{(2)}$ contributed by diagram
Fig.1(b.2) are
\begin{eqnarray}
\Lambda_{1}^{(2)}&=&\frac{1}{32\pi^2}\sum_{l=1}^4\sum_{\alpha=1}^2\left[\left(m_t
m_{\tilde{\chi}^0_l} (F_a^{(2)}+F_h^{(2)}) + m_t^2(
F_c^{(2)}+F_f^{(2)}) + (m_t^2  + m_t m_{\tilde{g}}
)(F_d^{(2)}+F_e^{(2)}) \right)C_{11}^{(2)}\right.\nb
\\&+&\left( (m_{\tilde{g}} -m_t) m_{\tilde{\chi}^0_l}( F_a^{(2)}+F_h^{(2)})
 + (m_{\tilde{g}}-m_t) m_t (F_c^{(2)}+F_f^{(2)}) +(m_{\tilde{g}}^2- m_t^2 )(F_d^{(2)}+F_e^{(2)})
 \right)C_{12}^{(2)}\nb
\\&+&\left(  m_{\tilde{g}} m_{\tilde{\chi}^0_l}
(F_a^{(2)}+F_h^{(2)}) + m_{\tilde{\chi}^0_l} m_t
(F_b^{(2)}+F_g^{(2)}) + m_t^2( F_c^{(2)}
+F_f^{(2)})\right.\nb\\&+&\left.
   ( m_t m_{\tilde{g}}+ m_{\tilde{t}_{\alpha}}^2
   )(F_d^{(2)}+F_e^{(2)})\right)C_0^{(2)}\nb
-\left.(F_d^{(2)}+F_e^{(2)})B_0^{(2)}\right]\nb\\
\Lambda_{2}^{(2)}&=&\frac{1}{32\pi^2}\sum_{l=1}^4\sum_{\alpha=1}^2\left[\left(m_t
m_{\tilde{\chi}^0_l} (F_a^{(2)}-F_h^{(2)}) + m_t^2(
F_c^{(2)}-F_f^{(2)})+
    (m_t m_{\tilde{g}}- m_t^2) (F_d^{(2)}-F_e^{(2)})\right)C_{11}^{(2)}\right.\nb
    \\&+& \left((m_t  + m_{\tilde{g}} )m_{\tilde{\chi}^0_l}(F_h^{(2)}- F_a^{(2)}) +(
    m_t + m_{\tilde{g}})m_t(F_f^{(2)}-F_c^{(2)}) + (m_t^2-
    m_{\tilde{g}}^2) (F_d^{(2)}-F_e^{(2)})\right)C_{12}^{(2)}\nb
    \\&+&\left(m_{\tilde{g}}
m_{\tilde{\chi}^0_l} (F_h^{(2)}-F_a^{(2)}) + m_{\tilde{\chi}^0_l}
m_t (F_g^{(2)}-F_b^{(2)}) + m_t^2(F_c^{(2)}-F_f^{(2)})
\right.\nb\\&+&\left. (m_t m_{\tilde{g}} -
m_{\tilde{t}_{\alpha}}^2) (F_d^{(2)}-F_e^{(2)})\right)C_0^{(2)}
+\left.\left(F_d^{(2)}-F_e^{(2)}\right)B_0^{(2)}\right]
\end{eqnarray}
with $B_0^{(2)} = B_0(-p1, m_{\tilde{\chi}_l^0},m_t)$,
$C_{0,11,12}^{(2)}=C_{0,11,12}(k_1, -p_1, m_{\tilde{t}_{\alpha}},
m_{\tilde{\chi}_l^0}, m_t).$
\item For diagram Fig.1(b.3), we introduce the following notation:
\begin{eqnarray}
F_a^{(3)}=V_{H^0_mtt}V_{H^0_m\tilde{t}_{\alpha}\tilde{t}_2}V_{\tilde{g}\tilde{t}_{\alpha}t}^{(1)},~~~~~~
F_b^{(3)}=V_{H^0_mtt}V_{H^0_m\tilde{t}_{\alpha}\tilde{t}_2}V_{\tilde{g}\tilde{t}_{\alpha}t}^{(2)}
\end{eqnarray}
The form factors $\Lambda_{1,2}^{(3)}$ contributed by diagram
Fig.1(b.3) are
\begin{eqnarray}
\Lambda_1^{(3)}=\frac{1}{32\pi^2}\sum_m^4\left[(F_a^{(3)}-F_b^{(3)})(m_t(C_0^{(3)}
+ C_{11}^{(3)}-C_{12}^{(3)})+m_{\tilde{g}}C_{12}^{(3)})\right]\nb \\
\Lambda_2^{(3)}=\frac{1}{32\pi^2}\sum_m^4\left[-(F_a^{(3)}+F_b^{(3)})
m_t(C_0^{(3)}+C_{11}^{(3)}-C_{12}^{(3)})+m_{\tilde{g}}C_{12}^{(3)})\right]
\end{eqnarray}
with $C_{0,11,12}^{(3)}=C_{0,11,12}(k_1, -p_1, m_t, m_{H_m^0},
m_{\tilde{t}_{\alpha}})$
\item For diagram Fig.1(b.4), we introduce the following notation:
\begin{eqnarray}
F_a^{(4)}=V_{H^+_ntb}^{(1)}V_{H^+_n\tilde{t}_2\tilde{b}_{\beta}}^{*}V_{\tilde{g}\tilde{b}_{\beta}b}^{(1)},~~~~~~
F_b^{(4)}=V_{H^+_ntb}^{(2)}V_{H^+_n\tilde{t}_2\tilde{b}_{\beta}}^{*}V_{\tilde{g}\tilde{b}_{\beta}b}^{(1)}\nb\\
F_c^{(4)}=V_{H^+_ntb}^{(1)}V_{H^+_n\tilde{t}_2\tilde{b}_{\beta}}^{*}V_{\tilde{g}\tilde{b}_{\beta}b}^{(2)},~~~~~~
F_d^{(4)}=V_{H^+_ntb}^{(2)}V_{H^+_n\tilde{t}_2\tilde{b}_{\beta}}^{*}V_{\tilde{g}\tilde{b}_{\beta}b}^{(2)}
\end{eqnarray}
The form factors $\Lambda_{1,2}^{(4)}$ contributed by diagram
Fig.1(b.4) are:
\begin{eqnarray}
\Lambda_1^{(4)}=\frac{1}{32\pi^2}\sum_n^2\left[m_b(F_a^{(4)}+F_d^{(4)})C_0^{(4)}
-m_t(F_b^{(4)}+F_c^{(4)})C_{11}^{(4)}+(m_t-m_{\tilde{g}})(F_b^{(4)}+F_c^{(4)})C_{12}^{(4)}\right]\nb \\
\Lambda_2^{(4)}=\frac{1}{32\pi^2}\sum_n^2\left[m_b(F_d^{(4)}-F_a^{(4)})C_0^{(4)}
-m_t(F_c^{(4)}-F_b^{(4)})C_{11}^{(4)}+(m_t-m_{\tilde{g}})(F_c^{(4)}-F_b^{(4)})C_{12}^{(4)}\right]\nb\\
~
\end{eqnarray}
with $C_{0,11,12}^{(4)}=C_{0,11,12}(k_1, -p_1, m_b, m_{H_n^+},
m_{\tilde{b}_{\beta}})$

\end{itemize}
\begin{flushleft} {\bf 4. Numerical results and conclusion } \end{flushleft}
\par
In the numerical analysis, we take the SM input parameters as:
$m_t=174.3$ GeV,$m_b=4.3$ GeV, $m_{Z}=91.1882$ GeV,
$m_{W}=80.419~GeV$ and $\alpha_{EW} = 1/128$ \cite{pdg}.
\par
Figure 2 shows the dependence on $m_{\tilde{g}}$ of the relative
correction $\delta \Gamma/\Gamma_0$ for the decay $\tilde{t}_2 \to
t+\tilde{g}$. For simplicity, we assumed
$M_{\tilde{Q}}=M_{\tilde{U}}=M_{\tilde{D}}=A_t=A_b= M_{SUSY}$. The
ratio of the vacuum expectation values $\tan\beta$ is set to be 4.
 The mass of charged Higgs boson
$m_{H^{\pm}} = 250$ GeV. The physical chargino masses
$m_{\tilde{\chi}_1^+}$, $m_{\tilde{\chi}_2^+}$ and the lightest
neutralino mass $m_{\tilde{\chi}_1^0}$ are set to be 100 GeV,
300 GeV and 60 GeV, respectively. Then the fundamental SUSY
parameters $M$, $M^{'}$ and $\mu$ in the chargino and neutralino
matrix can be extracted from these input chargino masses, lightest
neutralino mass $m_{\tilde{\chi}_1^0}$ and $\tan\beta$. The solid
curve is for $M_{SUSY}=400$ GeV and the dashed curve is for
$M_{SUSY}=500$ GeV. We can see the relative Yukawa correction to
this process is always negative when the $m_{\tilde{g}}$ is in the
range 80 to 300 GeV. The figure shows that the relative Yukawa
correction is not very sensitive to the value of $m_{\tilde{g}}$
and has the values about $-3\%$.
\par
In Fig. 3 we present the numerical result of the Yukawa correction
for the decay $\tilde{t}_2\to t+\tilde{g}$ in the minimal
supergravity (MSUGRA) scenario. The squark masses and mixing
parameters are calculated from the input MSUGRA parameters: the
common scalar mass $m_0$, the common gaugino mass $m_{1/2}$, the
trilinear coupling $A_0$, the ratio of the vacuum expectation
values of the Higgs fields $\tan\beta$, and the sign of the
Higgsino mass parameter $\mu$. Here we take $m_0=800$ GeV,
$A_0=200$ GeV, and $tan\beta=1.75$. The solid curve is for $\mu>0$
and the dashed curve is for $\mu<0$. In the case of $\mu<0$ and
$m_{1/2}>195$ GeV, the decay $\tilde{t}_2\to t+\tilde{g}$ cannot
be opened kinematically in the MSUGRA model. While in the case of
$\mu>0$ and $m_{1/2}>210$ GeV, the decay $\tilde{t}_2\to
t+\tilde{g}$ is forbidden. The figure shows that the relative
Yukawa correction to this process varies between $-9\%$ to $-2\%$
for $\mu>0$ and between $-6\%$ to $-3\%$ for $\mu<0$.
\par
Figure 4 shows the dependence on $m_{\tilde{g}}$ of the relative
correction for the decay $\tilde{g}\to \bar{t}+\tilde{t}_1$. Here
we choose $\tan\beta=4$, and $m_{H^{\pm}} = 250$ GeV. The physical
chargino masses $m_{\tilde{\chi}_1^+}$, $m_{\tilde{\chi}_2^+}$ and
the lightest neutralino mass $m_{\tilde{\chi}_1^0}$ are set to be
$100$ GeV, $300$ GeV and $60$ GeV, respectively. In Fig.4 the
solid curve is for $M_{SUSY}=200$ GeV and the dashed curve is for
$M_{SUSY}=150$ GeV. The relative Yukawa correction to this example
decreases from $1\%$ to $-4\%$ with the increasing of
$m_{\tilde{g}}$ from 400 to 800 GeV.
\par
In Fig.5 we present the numerical result of the Yukawa correction
for the decay $\tilde{g} \to \bar{t}+\tilde{t}_1$ in the mSUGRA
scenario. In this example we take $m_0=400$ GeV, $A_0=200$ GeV,
and $\tan\beta=1.75$. The solid curve and dashed curve are for
$\mu>0$ and $\mu<0$, respectively. The figure shows that when
$\mu<0$ and $m_{1/2}<237$ GeV, the decay $\tilde{g}\to
\bar{t}+\tilde{t}_1$ is not allowed kinematically, and when
$\mu>0$ and $m_{1/2}<145~GeV$, this decay is closed in the MSUGRA
scenario. We can see that in the case of $\mu>0 $, the relative
Yukawa correction is always negative and varies from $-5\%$ to
$-1\%$. For $\mu<0$, the relative correction decreases from $10\%$
to $-2\%$ with the increasing of $m_{1/2}$ from $145$ GeV to
$400$ GeV.
\par
In summary, we have computed the one-loop electroweak Yukawa
corrections to the partial widths of the $\tilde{t}_2\to
t+\tilde{g}$ and $\tilde{g}\to \bar{t}+\tilde{t}_1$ decays within
the minimal supersymmetric standard model. We find that the
relative corrections can be significant and we reach the value of
$10\%$ in some parameter space, which can be comparable with the
corresponding QCD corrections. Therefore, the electroweak
corrections to these decays of scalar top quark and gluino should
be taken into account in the precise experiment measurements.
\par
\noindent{\large\bf Acknowledgments:} This work was supported in
part by the National Natural Science Foundation of China and the
Education Ministry of China.

%\begin{flushleft} {\bf Appendix A} \end{flushleft}
\section{Appendix A}
\renewcommand{\theequation}{A.\arabic{equation}}
\par
 In this appendix, we list the self-energies of top squark and
top quark contributed by Fig.1(c) and Fig.1(d). The relevant
Feynman rules are presented in Refs.\cite{s3} \cite{sa}. We adopt
the notations of the couplings that chargino (neutralino) coupling
with quark and squark in Ref.\cite{zml}. They are written as
bellow
\begin{eqnarray}
\bar{b}-\tilde{t}_{i}-\bar{\tilde{\chi}}^+_j &:& \left(
V^{(1)}_{b\tilde{t}_i\tilde{\chi}^+_j}P_L +
 V^{(2)}_{b\tilde{t}_i\tilde{\chi}^+_j}P_R\right)C,~~~
\bar{t}-\tilde{b}_{i}-\tilde{\chi}^+_j:V^{(1)}_{t\tilde{b}_i\tilde{\chi}^+_j}P_L
+
V^{(2)}_{t\tilde{b}_i\tilde{\chi}^+_j}P_R, \nb\\
\bar{b}-\tilde{b}_{i}-\tilde{\chi}^0_j&:&V^{(1)}_{b\tilde{b}_i\tilde{\chi}^0_j}P_L
+ V^{(2)}_{b\tilde{b}_i\tilde{\chi}^0_j}P_R,~~~
\bar{t}-\tilde{t}_{i}-\tilde{\chi}^0_j:V^{(1)}_{t\tilde{t}_i\tilde{\chi}^0_j}P_L
+ V^{(2)}_{t\tilde{t}_i\tilde{\chi}^0_j}P_R,
\end{eqnarray}
where $C$ is the charge conjugation operator.
\par
Defining $H^0_m=(h^0,H^0,A^0,G^0)(m=1,2,3,4)$ and
$H^{\pm}_n=(H^{\pm},G^{\pm})(n=1,2)$, the couplings between
$H^0_m(H^{\pm}_n)$ and quark(squark) are denoted as
\begin{eqnarray}
H^+_n-\bar{t}-b &:&~ V^{(1)}_{H^+_ntb}P_L +
V^{(2)}_{H^+_ntb}P_R,~~~~~H^+_n-\tilde{t}_i-\tilde{b}_j:~
V_{H^+_n\tilde{t}_i\tilde{b}_j}, \nb\\
H^+_n-H^-_n-\tilde{t}_i-\tilde{t}_j&:&~V_{H^+_nH^-_n\tilde{t}_i\tilde{t}_j},\nb\\
H^0_m-\bar{t}-t &:&~ V_{H^0_mtt}~~(for~~m=1,2);~~~ \gamma_5V_{H^0_mtt}~~(for~~ m=3,4),\nb\\
H^0_m-\tilde{t}_i-\tilde{t}_j&:&~
V_{H^0_m\tilde{t}_i\tilde{t}_j},~~~~~H^0_m-H^0_m-\tilde{t}_i-\tilde{t}_j:~V_{H^0_mH^0_m\tilde{t}_i\tilde{t}_j}.
\end{eqnarray}
\par
The couplings between four squarks are denoted as
\begin{eqnarray}
\tilde{t}_i-\tilde{t}_j-\tilde{t}_k-\tilde{t}_l:~V_{\tilde{t}_i\tilde{t}_j\tilde{t}_k\tilde{t}_l}~~~~~~
\tilde{t}_i-\tilde{t}_j-\tilde{b}_k-\tilde{b}_l:~V_{\tilde{t}_i\tilde{t}_j\tilde{b}_k\tilde{b}_l}
\end{eqnarray}
\par And we denote the couplings between gluino , quark and
squark as
\begin{eqnarray}
\tilde{g}-\tilde{t}_i-\bar{t}:~V^{(1)}_{\tilde{g}\tilde{t}_it}P_L
+V^{(2)}_{\tilde{g}\tilde{t}_it}P_R~~~~~~\tilde{g}-\tilde{b}_i-\bar{b}:~V^{(1)}_{\tilde{g}\tilde{b}_ib}P_L
+V^{(2)}_{\tilde{g}\tilde{b}_ib}P_R
\end{eqnarray}
\par
The stop quark self-energy contributed by Fig.1(c1)-(c3) reads
\begin{eqnarray}
\Sigma_{ij}^{\tilde{t}}(p^2)&=&\frac{1}{16\pi^2}\{\sum_m^4\sum_{\alpha}^2\left[-2iA_0(m_{H^0_m})V_{H^0_mH^0_m\tilde{t}_i\tilde{t}_j}-B_0(-p,
m_{\tilde{t_{\alpha}}},m_{H^0_m})V_{H^0_m\tilde{t}_j\tilde{t}_{\alpha}}V_{H^0_m\tilde{t}_{\alpha}\tilde{t}_i}\right]\nb \\
&-&\sum_n^2\sum_{\beta}^2\left[iA_0(m_{H^+_n})V_{H^+_nH^-_n\tilde{t}_i\tilde{t}_j}+B_0(-p,
m_{\tilde{b_{\beta}}},m_{H^+_n})V_{H^+_n\tilde{t}_i\tilde{b}_{\beta}}V_{H^+_n\tilde{t}_j\tilde{b}_{\beta}}\right]\nb\\
&-&\sum_{l}^4\left[2\left(V_{t\tilde{t}_j\tilde{\chi}^0_l}^{(1)*}V_{t\tilde{t}_i\tilde{\chi}^0_l}^{(1)}
+V_{t\tilde{t}_j\tilde{\chi}^0_l}^{(2)*}V_{t\tilde{t}_i\tilde{\chi}^0_l}^{(2)}\right)
\left(A_0(m_{\tilde{\chi}^0_l})+B_0(-p,m_t,m_{\tilde{\chi}^0_l})m_t^2+B_1(-p,m_t,m_{\tilde{\chi}^0_l})p^2\right)\right.\nb\\
&+&\left.2B_0(-p,m_t,m_{\tilde{\chi}^0_l})m_tm_{\tilde{\chi}_l^0}
\left(V_{t\tilde{t}_j\tilde{\chi}^0_l}^{(2)*}V_{t\tilde{t}_i\tilde{\chi}^0_l}^{(1)}
+V_{t\tilde{t}_j\tilde{\chi}^0_l}^{(1)*}V_{t\tilde{t}_i\tilde{\chi}^0_l}^{(2)}\right)\right]\nb\\
&-&\sum_k^2\left[2\left(V_{b\tilde{t}_j\tilde{\chi}^+_k}^{(1)*}V_{b\tilde{t}_i\tilde{\chi}^+_k}^{(1)}
+V_{b\tilde{t}_j\tilde{\chi}^+_k}^{(2)*}V_{b\tilde{t}_i\tilde{\chi}^+_k}^{(2)}\right)
\left(A_0(m_{\tilde{\chi}^+_k})+B_0(-p,m_b,m_{\tilde{\chi}^+_k})m_b^2+B_1(-p,m_b,m_{\tilde{\chi}^+_k})p^2\right)\right.\nb\\
&+&\left.2B_0(-p,m_b,m_{\tilde{\chi}^+_k})m_bm_{\tilde{\chi}_k^+}
\left(V_{b\tilde{t}_j\tilde{\chi}^+_k}^{(2)*}V_{b\tilde{t}_i\tilde{\chi}^+_k}^{(1)}
+V_{b\tilde{t}_j\tilde{\chi}^+_k}^{(1)*}V_{b\tilde{t}_i\tilde{\chi}^+_k}^{(2)}\right)\right]\nb\\
&-&\sum_{\alpha,\beta}^2\left[iA_0(m_{\tilde{t}_{\alpha}})V_{\tilde{t}_i\tilde{t}_{\alpha}\tilde{t}_{\alpha}\tilde{t}_j}
  +iA_0(m_{\tilde{b}_{\beta}})V_{\tilde{t}_i\tilde{b}_{\beta}\tilde{b}_{\beta}\tilde{t}_j}\right]\}
\end{eqnarray}
\par
The top quark self-energy contributed by Fig.1(d1) and 1(d2) reads
\begin{eqnarray}
\Sigma_{tt}(p^2) &=& i (\rlap/{p} \Sigma_{tt}^L P_L
+\rlap/{p}\Sigma_{tt}^R P_R+\Sigma_{tt}^{SL}
P_L+\Sigma_{tt}^{SR}P_R),
\end{eqnarray}
where
\begin{eqnarray}
\Sigma_{tt}^L &=&\frac{1}{16 \pi^2}\{-\sum_{m}^4[(V_{H^0_mtt})^2
B_1(-p,m_t,m_{H^0_m})] - \sum_n^2[(V_{H^+_ntb}^{(2)})^2B_1(-p,m_b,m_{H^+_n})]\nb\\
&+&\sum_k^2\sum_{\beta}^2[(V_{t\tilde{b}_{\beta}\tilde{\chi}^+_k}^{(2)})^2(B_0+B_1)(-p,m_{\tilde{b}_{\beta}},m_{\tilde{\chi}^+_k})]
+\sum_l^4\sum_{\alpha}^2[(V_{t\tilde{t}_{\alpha}\tilde{\chi}^0_l}^{(2)})^2(B_0+B_1)(-p,m_{\tilde{t}_{\alpha}},m_{\tilde{\chi}^0_l})]\}\nb\\
\Sigma_{tt}^R &=& \Sigma_{tt}^L\left(V_{H^+_ntb}^{(2)}\to
V_{H^+_ntb}^{(1)},V_{t\tilde{b}_{\beta}\tilde{\chi}^+_k}^{(2)}\to
V_{t\tilde{b}_{\beta}\tilde{\chi}^+_k}^{(1)},V_{t\tilde{t}_{\alpha}\tilde{\chi}^0_l}^{(2)}\to
V_{t\tilde{t}_{\alpha}\tilde{\chi}^0_l}^{(1)}\right)\nb\\
\Sigma_{tt}^{SL}&=&\frac{1}{16\pi^2}\{-\sum_{m}^4[(V_{H^0_mtt})^2m_t
B_1(-p,m_t,m_{H^0_m})]+\sum_n^2[V_{H^+_ntb}^{(2)*}V_{H^+_ntb}^{(1)}m_bB_1(-p,m_b,m_{H^+_n})]\nb\\
&+&\sum_k^2\sum_{\beta}^2[V_{t\tilde{b}_{\beta}\tilde{\chi}^+_k}^{(2)*}V_{t\tilde{b}_{\beta}\tilde{\chi}^+_k}^{(1)}
m_{\tilde{\chi}^+_k}B_0(-p,m_{\tilde{b}_{\beta}},m_{\tilde{\chi}^+_k})]
+\sum_l^4\sum_{\alpha}^2[V_{t\tilde{t}_{\alpha}\tilde{\chi}^0_l}^{(2)*}V_{t\tilde{t}_{\alpha}\tilde{\chi}^0_l}^{(1)}
m_{\tilde{\chi}^0_l}B_0(-p,m_{\tilde{t}_{\alpha}},m_{\tilde{\chi}^0_l})]\}\nb\\
\Sigma_{tt}^{SR}&=&\Sigma_{tt}^{SL}\left(
V_{H^+_ntb}^{(2)*}V_{H^+_ntb}^{(1)}\to
V_{H^+_ntb}^{(1)*}V_{H^+_ntb}^{(2)},
V_{t\tilde{b}_{\beta}\tilde{\chi}^+_k}^{(2)*}V_{t\tilde{b}_{\beta}\tilde{\chi}^+_k}^{(1)}\to
V_{t\tilde{b}_{\beta}\tilde{\chi}^+_k}^{(1)*}V_{t\tilde{b}_{\beta}\tilde{\chi}^+_k}^{(2)},
V_{t\tilde{t}_{\alpha}\tilde{\chi}^0_l}^{(2)*}V_{t\tilde{t}_{\alpha}\tilde{\chi}^0_l}^{(1)}\to
V_{t\tilde{t}_{\alpha}\tilde{\chi}^0_l}^{(1)*}V_{t\tilde{t}_{\alpha}\tilde{\chi}^0_l}^{(2)}\right)\nb\\
~~~~~
\end{eqnarray}
\par
In our paper we adopt the definitions of the one-loop integrals in
the Ref. \cite{s13}. The numerical calculation of the vector
and tensor one-loop integral functions can be traced back to four
scalar loop integrals $A_{0}$, $B_{0}$, $C_{0}$, $D_{0}$ as shown
in \cite{passvelt}.

\vskip 10mm
\begin{flushleft} {\bf Figure Captions} \end{flushleft}

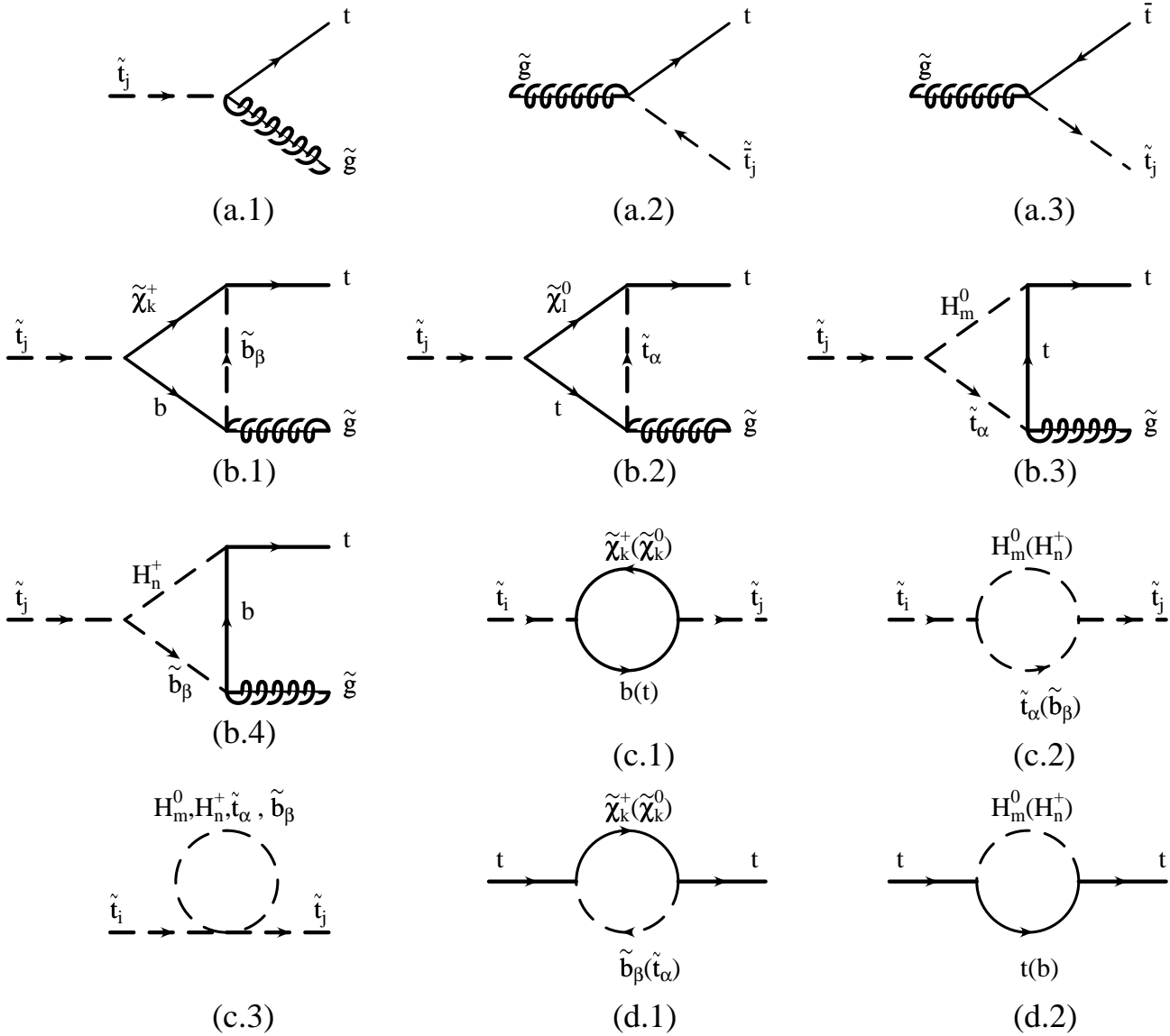
{\bf Fig.1} Feynman diagrams including one-loop Yukawa corrections
to the decays $\tilde{t}_j \to \tilde{g}+t$ and $\tilde{g}\to
\bar{t}+\tilde{t}_j$: Fig.1(a) tree-level diagram. Fig.1(b) vertex
corrections for stop decays. Fig.1(c) stop quark self-energies.
Fig.1(d) top quark self-energies. In these Figures
$H^0_{m}=h^0,H^0,A^0,G^0(m=1-4)$ and $H^+_{n}=H^+,G^+(n=1,2)$.
\par
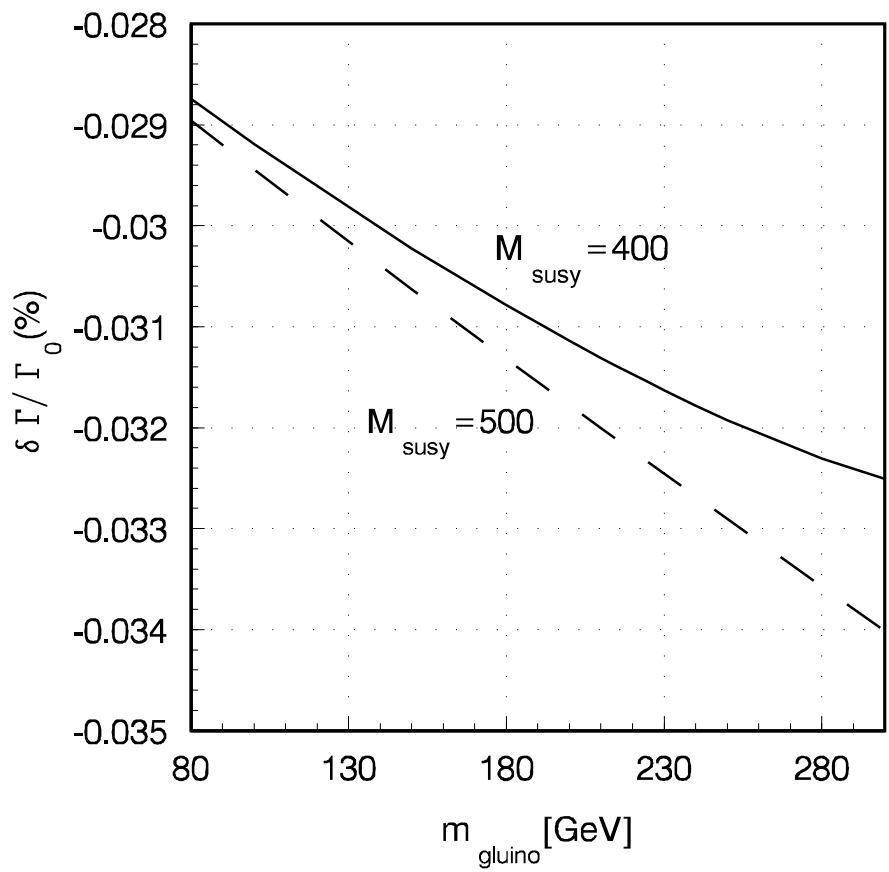
{\bf Fig.2} The relative correction for $\tilde{t}_2\to
\tilde{g}+t$ as a function of $m_{\tilde{g}}$.
\par
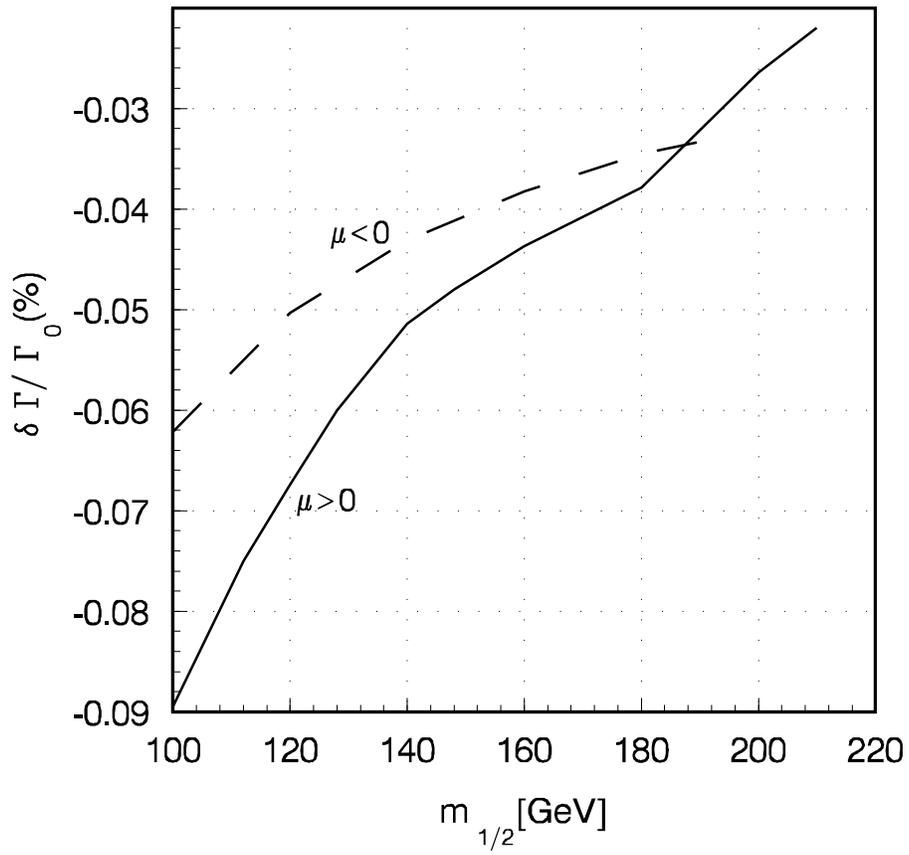
{\bf Fig.3} The relative correction for $\tilde{t}_2\to
\tilde{g}+t$ as a function of $m_{1/2}$ in the mSUGRA scenario.
\par
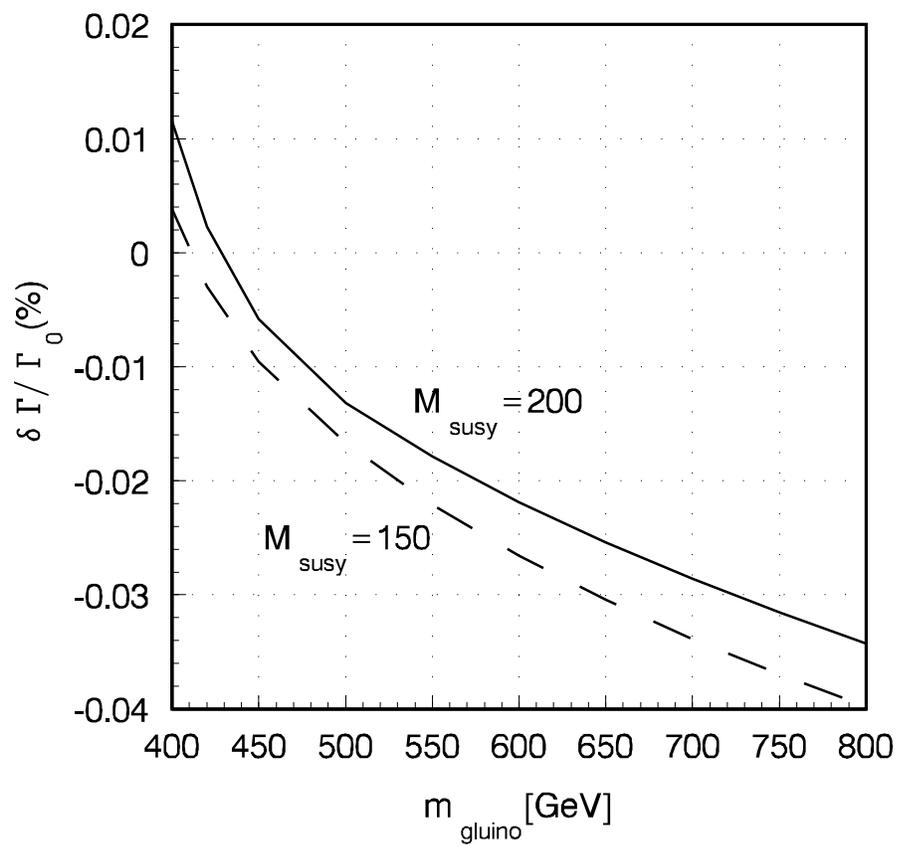
{\bf Fig.4} The relative correction for $\tilde{g}\to
\bar{t}+\tilde{t}_1$ as a function of $m_{\tilde{g}}$.
\par
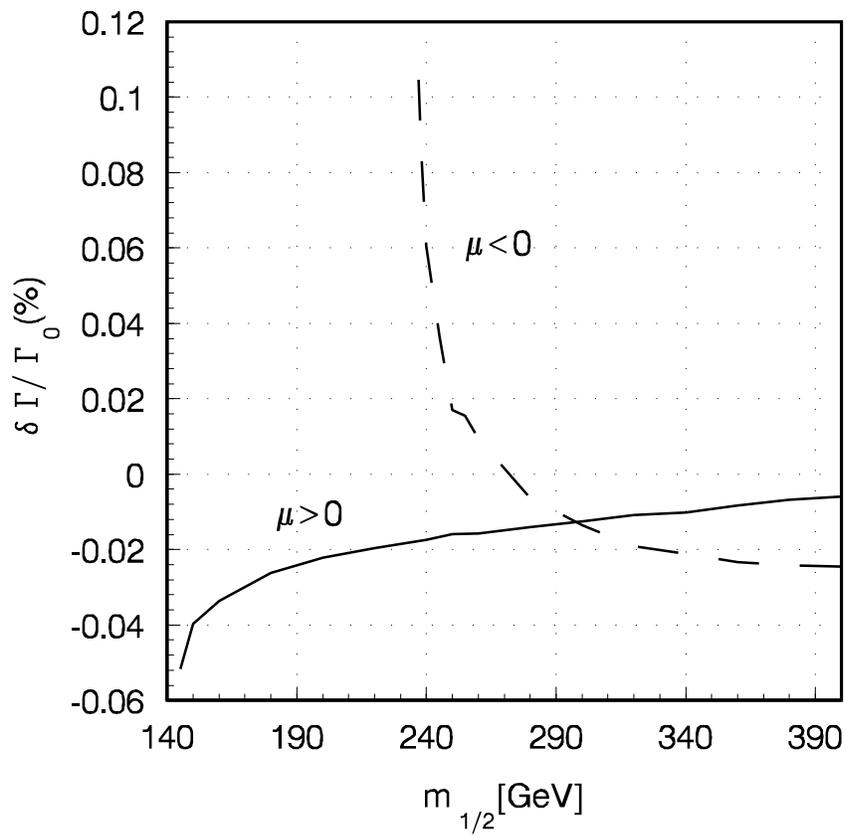
{\bf Fig.5} The relative correction for $\tilde{g}\to
\bar{t}+\tilde{t}_1$ as a function of $m_{1/2}$ in the mSUGRA
scenario.
\end{document}